\documentclass[%
 reprint,
preprintnumbers,
nofootinbib,
 amsmath,amssymb,
 prd,
]{revtex4-2}

\usepackage{graphicx}
\usepackage{dcolumn}
\usepackage{bm}
\usepackage{booktabs}
\usepackage{xcolor}
\usepackage{slashed}
\usepackage{amssymb}
\usepackage{multirow}

\usepackage[hidelinks]{hyperref}

\newcommand{\mth}{m_\mathrm{th}}
\newcommand{\ms}{m_s}
\newcommand{\sintwotheta}{\sin^2(2\theta)}
\newcommand{\gV}{g_V}
\newcommand{\mV}{m_V}

\begin{document}

\preprint{UCI-TR-2025-12, INT-PUB-25-024}

\title{Cosmic Structure Strikes Back: The Elimination of Vector–Mediated Nonstandard Interaction Models as a Mechanism for Sterile Neutrino Dark Matter Production}

\author{Cannon M.\ Vogel}
\email{cvogel1@uci.edu}
\affiliation{Center for Cosmology, Department of Physics and Astronomy, University of California---Irvine, Irvine, CA 92697-4575, USA}

\author{Helena Garc\'ia Escudero}
\email{garciaeh@uci.edu}
\affiliation{Center for Cosmology, Department of Physics and Astronomy, University of California---Irvine, Irvine, CA 92697-4575, USA}

\author{Kevork N.\ Abazajian}
\email{kevork@uci.edu}
\affiliation{Center for Cosmology, Department of Physics and Astronomy, University of California---Irvine, Irvine, CA 92697-4575, USA}

\date{\today}

\begin{abstract}
We revisit sterile-neutrino production enabled by nonstandard interactions (NSI) among active neutrinos mediated by new bosons. We focus on vector mediators, including neutrinophilic, gauged $L_\mu\!-\!L_\tau$, and $B\!-\!L$ realizations that modify in-medium dispersion and scattering, thereby altering the active--sterile conversion history. Building on a novel production framework with NSI thermal potentials and collision integrals, we compute nonthermal phase-space distributions across sterile neutrino mixing and NSI parameters and map each point to an \emph{equivalent thermal warm dark matter} particle mass $\mth$ via linear-theory transfer function fitting with cosmological structure-formation Boltzmann solver. This enables a direct reinterpretation of state-of-the-art structure-formation limits from Milky Way satellites, strong lensing, and the Lyman-$\alpha$ forest. These limits, in conjunction with X-ray decay searches, as well as results from a wide variety of particle physics experiments allow for a more complete examination of these models. We find that these vector-mediated models are ruled out when the full combination of current constraints, listed above, are taken into account. 
NSI scalar-mediated models and models with low-reheating temperatures remain viable. 
\end{abstract}

\maketitle

\section{Introduction}\label{sec:intro}

The particle identity of the dark matter (DM) is unknown, but sterile neutrinos remain a particularly well-motivated candidate, emerging naturally in neutrino-mass models and offering multi-pronged probes via nuclear decay searches, as well as in cosmology and X-ray astronomy~\cite{Kusenko:2009up,Abazajian:2017tcc}. Sterile neutrinos are an electroweak-singlet state, $\nu_s$, that can mix with $\nu_\alpha$ of the Standard Model flavors $\nu_e,\nu_\mu,\nu_\tau$. The lightest sterile flavor eigenstate can be written as
\begin{equation}\label{eq:Sterile}
\nu_s = \cos\theta \; \nu_4 + \sin\theta \; \nu_i ,
\end{equation}
where $\nu_4$ is the mass eigenstate most closely associated with the sterile neutrinos, and $\nu_i$ is the lighter mass eigenstate with which the sterile mixes in the two-neutrino mixing approximation. Given the requirement of the small-mixing $\theta \ll 1$ for mechanisms of sterile neutrino dark matter production, we will follow standard usage and refer to mass of $\nu_4$ as that of the sterile neutrino.

Production via oscillations in the early Universe was first proposed by Dodelson and Widrow (DW) \cite{Dodelson:1993je} and later extended to resonant conversion in the presence of lepton asymmetry by Shi and Fuller (SF) \cite{Shi:1998km}. 
The DW mechanism---involving only Standard Model active neutrino interactions---is now excluded as the sole source of the full dark matter density, based on a combination of X-ray non-detections and structure-formation constraints \cite{Horiuchi:2013noa}. In contrast, SF production remains viable within regions of $(m_s,\sin^2 2\theta)$ where small mixing angles and large lepton asymmetries provide sufficient freedom (see, e.g., Refs.~\cite{Venumadhav:2015pla,Cherry:2017dwu,Akita:2025txo}), particularly when accounting for recent results that permit large lepton asymmetries during production, as discussed in the companion paper Ref.~\cite{Vogel:2025aut}.

A different and distinct avenue for dark matter production leverages enhanced active-neutrino interactions via nonstandard interactions (NSI) specifically among active neutrinos mediated by new bosons which are not part of the Standard Model. These augmented active-neutrino interactions increase sterile neutrino production for a given mixing angle in oscillation-based production. The first models considered for augmenting the active neutrino interactions involved a new scalar \cite{DeGouvea:2019wpf}, and subsequent models considered vector mediators for the self-interaction \cite{Kelly:2020pcy,Chichiri:2021wvw}. These new interaction cases have distinct laboratory signatures as well, which complement cosmological sensitivities to these properties. Such interactions modify the in-medium potential and augment collisional damping that govern active--sterile conversion, allowing the DW mechanism to potentially remain viable in producing the full dark matter density when including appropriate levels of these interactions. These NSI works have emphasized that the combination of enhanced scattering and altered dispersion relations can yield sterile-neutrino DM with phase-space distributions (PSDs) and linear power spectra that differ from both standard DW and SF cases. Moreover, structure-formation data have begun to constrain specific NSI production scenarios. For example, strong self-coupling with heavy ($\gtrsim$GeV) scalar mediators are now excluded as an origin for \emph{all} of the DM~\cite{An:2023mkf}. 

An alternative series of production mechanisms consider new bosons mediating sterile neutrino interactions, including a new vector $Z^\prime$ which can produce the sterile neutrino dark matter via decay \cite{Shuve:2014doa}. 
Or, the mediator can modify sterile neutrino self-interactions, altering the Boltzmann production process \cite{Johns:2019cwc}, and can open up a new parameter space for sterile neutrino dark matter production \cite{Bringmann:2022aim,Astros:2023xhe,Dev:2025sah}. In another mechanism, the mediator can transfer oscillation-based sterile neutrinos to lower-mass sterile neutrinos for dark matter production \cite{Fuller:2024noz}.

In this work, we present a systematic study of sterile neutrino dark matter produced through \emph{vector}-mediated NSI. Our central method is to map each model point onto an equivalent thermal warm dark matter (WDM) particle mass, $\mth$, using linear-theory transfer functions determined from linear structure formation Boltzmann calculations. This approach enables the direct application of the latest cosmological data, including constraints from Milky Way satellite counts and strong-lensing substructure, which together require $\mth > 8.5$~keV (95\% ~C.L.)~\cite{Zelko:2022tgf},\footnote{We have modified these $m_\mathrm{th}$ values to the more accurate thermal WDM transfer functions of Ref.~\cite{Vogel:2022odl}, and provide this correction in Eq.~7 in Ref.~\cite{Vogel:2025aut}. Within their respective analyses, using prior thermal WDM transfer functions, these works infer a thermal WDM preference of $m_\mathrm{th} = 4.5^{+40.5}_{-1.4}\,\mathrm{keV}$ in the Lyman-$\alpha$ forest data \cite{Villasenor:2022aiy} and constraints of $m_\mathrm{th} > 9.7\,\mathrm{keV}$ (95\% CL) \cite{Nadler:2021dft} for galaxy counts,   $m_\mathrm{th} > 9.8\,\mathrm{keV}$ (95\% CL) from a combination of galaxy counts and strong lensing \cite{Zelko:2022tgf}.}  as well as the preference for WDM from the Lyman-$\alpha$ forest flux power spectrum, $\mth = 4.1^{+33.3}_{-1.5}$~keV~\cite{Villasenor:2022aiy}. Sterile neutrino dark matter, however, is \textit{nonthermal}, and these limits must be converted to where its particle mass produces commensurate effects \cite{Zelko:2022tgf}. 

Complementary bounds on sterile neutrino dark matter arise from X-ray searches targeting the monochromatic decay line at $E_\gamma \simeq \ms/2$ \cite{Abazajian:2001nj,Abazajian:2001vt}. These searches provide stringent constraints on the active-sterile mixing angle, $\sintwotheta$, across a wide mass range, with leading limits established by Chandra, XMM-Newton, NuSTAR, and INTEGRAL/SPI in the $\sim$5--1000~keV window~\cite{Sicilian:2020glg, Ruchayskiy:2015onc, Jeltema:2015mee, Roach:2022lgo, Krivonos:2024yvm, Fischer:2022pse}. There has been considerable interest at the 7 keV particle-mass scale due to the detection of an unidentified line at 3.5 keV \cite{Bulbul:2014sua,Boyarsky:2014jta}, though this signal is highly constrained by the aforementioned limits.

Our goals are threefold: (i) present a transparent production and structure formation analysis framework for several vector NSI models that handles a large range of parameter values between the light and heavy mediator regimes; (ii) quantify, via the $\mth$ mapping, how small-scale structure and astronomical X-ray observations discriminate among NSI models and their internal parameter spaces; and (iii) combine the former constraints with those from the particular particle physics models, arising both from astrophysical and lab-based searches, in order to develop a complete picture of the current state of vector-mediated NSI-based production.

\section{Vector-Mediated Nonstandard Interaction Models}
\label{sec:NSI_models}

Following the framework of Ref.~\cite{Kelly:2020pcy}, we consider three classes of vector-mediated nonstandard interaction (NSI) models for sterile neutrino dark matter. In each case, the sterile state $\nu_s$ mixes slightly with an active neutrino, which we take to be the muon neutrino, enabling production via oscillations in the early Universe. The novel ingredient is an additional vector boson $V$, which modifies both the active-neutrino thermal potential $V_T$ and scattering rate $\Gamma$, reshaping the production history compared to the standard DW scenario. Below we summarize the essential features of each model.

\subsection{Neutrinophilic Vector}
\label{sec:neutrinophilic_summary}

The simplest possibility is a vector boson that couples only to active neutrinos. At low energies, such couplings can be written as
\begin{equation}
\mathcal{L}_{\nu V} \;=\; \sum_{\alpha,\beta=e,\mu,\tau} \lambda_{\alpha\beta}\, 
\overline{\nu}_\alpha \gamma^\rho \nu_\beta \, V_\rho \, ,
\end{equation}
with effective strengths $\lambda_{\alpha\beta}$ arising from higher-dimensional operators.  

This ``secret'' interaction enhances $\nu$--$\nu$ scattering, prolongs thermalization, and modifies $V_T$ such that sterile neutrinos can be produced efficiently through active–sterile oscillations. Depending on the mediator mass $m_V$ and coupling strength, three distinct regimes emerge: production dominated by heavy off-shell exchange, by on-shell vector decays, or by light-mediator scattering (see~\cite{Kelly:2020pcy} for details). 
\subsection{Gauged $U(1)_{L_\mu-L_\tau}$}
\label{sec:LmuLtau_summary}

A second realization is the anomaly-free $U(1)_{L_\mu-L_\tau}$ gauge extension. The new boson couples to second- and third-generation leptons as
\begin{align}
\mathcal{L}_{L_\mu-L_\tau} \;=\;& g_{\mu\tau} V_\alpha \big(
 \overline{\mu}\gamma^\alpha \mu - \overline{\tau}\gamma^\alpha \tau \cr
& + \overline{\nu}_\mu \gamma^\alpha P_L \nu_\mu  - \overline{\nu}_\tau \gamma^\alpha P_L \nu_\tau \big)\, .
\end{align}
As in the neutrinophilic scenario, we assume that the sterile neutrino dark matter state $\nu_4$ contains a small $\nu_\mu$ component, parametrized by a mixing angle $\theta$. However, this mixing cannot arise directly from the Yukawa term in Eq.~(\ref{eq:Sterile}), since $\nu_s$ must remain a singlet under $U(1)_{L_\mu-L_\tau}$ to prevent it from thermalizing fully in the early Universe. Instead, the mixing is generated via a higher-dimensional operator,  
\begin{equation}\label{eq:Sterile2}
y_\varepsilon \, \nu_s^T C H^T (i\sigma_2) L \left(\frac{\phi}{\Lambda}\right) + {\rm h.c.} \ ,
\end{equation}
where $\phi$ is a complex scalar field that Higgses the $U(1)_{L_\mu-L_\tau}$ symmetry and carries charge opposite to that of $L=(\nu_\mu,\mu)^T$. This construction ensures that $\nu_s$ mixes only weakly with $\nu_\mu$ while avoiding full equilibrium with the thermal bath, thereby preserving the conditions for oscillation-based production \cite{Kelly:2020pcy}.

In this case, sterile neutrino production again proceeds through oscillations, but now driven by $\nu_\mu$ or $\nu_\tau$ scattering mediated by $V$. The mechanism resembles the neutrinophilic scenario, but with additional charged-lepton couplings that alter both early-Universe scattering rates and laboratory phenomenology. This model is notable for also potentially addressing the $(g-2)_\mu$ anomaly~\cite{Baek:2001kca, Harigaya:2013twa}, providing complementary motivation beyond dark matter. As shown in Ref.~\cite{Kelly:2020pcy}, the same trio of production regimes is present, but with a more complex set of interaction rates due to the leptonic physics.

\subsection{Gauged $U(1)_{B-L}$}
\label{sec:BL_summary}

Finally, we consider the gauged $U(1)_{B-L}$ symmetry, a well-motivated extension of the Standard Model in which all fermions carry a universal charge. The corresponding interaction is
\begin{equation}
\mathcal{L}_{B-L} \;=\; g_{BL} V_\alpha \left( \tfrac{1}{3}\bar{q}\gamma^\alpha q 
- \bar{\ell}\gamma^\alpha \ell - \bar{\nu}\gamma^\alpha P_L \nu \right),
\end{equation}
where $V_\alpha$ is the new gauge boson.  

In the early Universe, this coupling modifies both the thermal potential and scattering rates of active neutrinos, allowing for resonant production of sterile neutrino dark matter. Compared with the flavor-specific $U(1)_{L_\mu-L_\tau}$ case, however, the universal coupling to electrons and quarks leads to much stronger laboratory and collider bounds. As a result, the viable parameter space for achieving the observed relic abundance is significantly more restricted.  

Current exclusions from beam dumps, $e^+e^-$ colliders, and precision experiments already rule out much of the region compatible with keV-scale sterile neutrino dark matter. The remaining window will be fully probed in the near future by Belle-II \cite{Belle-II:2018jsg}, LDMX \cite{Berlin:2018bsc}, and the LHC forward experiment FASER \cite{Feng:2017uoz}. Thus, while $U(1)_{B-L}$ provides a natural and predictive framework for vector-mediated nonstandard neutrino interactions, it is also the most experimentally constrained of the models considered here.

\bigskip
In summary, these three models illustrate distinct but related incarnations of vector-mediated NSIs. Each modifies active neutrino dynamics in the early Universe, reshaping sterile neutrino dark matter production and providing experimentally testable signatures. In the following, we extend the analysis of Ref.~\cite{Kelly:2020pcy} by connecting these models directly to thermal WDM equivalents and the corresponding structure-formation constraints, before mapping directly onto the full sterile neutrino parameter space.

\section{Methods}\label{sec:methods}

\subsection{Production with NSI}
\label{subsec:production}

As in the DW mechanism, sterile neutrino dark matter is produced through active--sterile oscillations in the thermal bath. The crucial difference is that the new vector boson modifies both the thermal potential experienced by active neutrinos as well as their collisional rate, and thus damping. These modifications can qualitatively alter the production history.

The evolution of the sterile phase-space distribution $f_s(p,T)$ is governed by a Boltzmann-like equation, in which the rate of production per mode depends on the effective in-medium mixing angle. In schematic form,
\begin{align}
\frac{df_s}{d \ln{z}} &\;=\; \frac{\Gamma_\alpha}{4H}\,\sin^2 2\theta_\mathrm{eff}\,
   \; f_\alpha , \nonumber\\[6pt]
\sin^2 2\theta_\mathrm{eff} &\;=\;
   \frac{\sin^2 2\theta}{\sin^2 2\theta+(\cos 2\theta - \mathcal{V})^2 + \mathcal{D}^2},
\end{align}
where $\mathcal{V} \equiv 2EV_T/\Delta m^2$ encodes matter effects from forward scattering in the medium, and $\mathcal{D} \equiv \Gamma_\alpha E/\Delta m^2$ represents collisional damping, sometimes referred to as the quantum Zeno effect. The term $\Delta m^2=m_s^2-m_a^2 \approx m_s^2$, for keV-scale sterile neutrinos. The variable $z\equiv \mu/T$, with $\mu \equiv 1 \; \mathrm{MeV}$ is used to conveniently parameterize cosmic time. 

The opacity $\Gamma_\alpha$ and potential $V_T$ now include NSI contributions from $V$ exchange in both forward scattering and $2\leftrightarrow 2$ processes. These quantities are evaluated with the appropriate thermal corrections: finite-temperature screening of the propagator, instantaneous decoupling of species as they become nonrelativistic, and the opening or closing of additional scattering channels. In practice, the structure of the production equation remains the same as in the Standard Model case, but the parametric dependence of $V_T$ and $\Gamma_\alpha$ on the new coupling $g_V$ and mediator mass $m_V$ can change production dramatically.

This framework smoothly interpolates using precomputed integrals between the known analytic limits. In the heavy-mediator regime ($T \ll m_V$), the NSI reduce to contact interactions, reproducing the scaling $\Gamma \propto g_V^4 T^5/m_V^4$ and a negative thermal potential $\propto -g_V^2 E T^4/m_V^4$, in agreement with~\cite{Kelly:2020pcy}. In the opposite light-mediator limit ($T \gg m_V$), the mediator behaves like a new relativistic species in the plasma, yielding a positive contribution to the potential $\propto g_V^2 T^2/E$, again consistent with analytic expectations. We have benchmarked our numerical implementation against these limits, and against the scalar-mediated case of~\cite{DeGouvea:2019wpf}, which is recovered upon the replacements $g_V \to \lambda$ and $m_V \to m_\phi$.

In summary, NSI leave the formal structure of the production mechanism intact but substantially reshape the thermal environment in which oscillations occur. Depending on the sign and magnitude of the induced potential, production can be resonantly enhanced, suppressed until late times, or dominated by mediator decay. This richer phenomenology underpins the distinct parameter space features we discuss in the following sections.

\paragraph*{Numerics and tables.}
For speed and stability we precompute a grid of NSI potentials and collision integrals on a table in $(E/T,\ T,\ \mV/T,\ \gV)$ spanning the ranges used in the scans, with logarithmic spacing in all ratios. We verify table interpolation against direct integrals at random points to percent-level for $V_T$ and few-percent for $\Gamma_\alpha$ in both mediator regimes.

\paragraph*{Grids and parameters.}
We perform logarithmic grids in the model parameters:
\begin{align}
\ms/\mathrm{keV}&\in [10^0,10^3],\qquad
\sintwotheta\in[10^{-23},10^{-6}],\nonumber\\
\gV&\in[10^{-8},10^{0}],\qquad
\mV/\mathrm{GeV}\in[10^{-3},10^{2}],
\label{eq:grids}
\end{align}
with fixed mediator choices per model (neutrinophilic, $L_\mu\!-\!L_\tau$, $B\!-\!L$). For scalar comparisons we use the same $\ms$--$\sintwotheta$ grids with representative $m_\phi$ benchmarks as in~\cite{DeGouvea:2019wpf, An:2023mkf}. Throughout this work, we integrate the driving equation from $T_{\mathrm{high}}\approx 100 \; \mathrm{GeV}$ to $T_{\mathrm{low}}\approx 100 \; \mathrm{keV}$ in order to capture all possible production, although in many cases relevant production only occurs in a small window of cosmic history. 

\subsection{Relating $g_V$ and $m_V$}

For a sterile neutrino of given $(m_s,\sin^22\theta)$, successful production of the observed dark matter density does not occur uniformly across all values of the mediator coupling $g_V$ and mass $m_V$. Instead, only specific combinations of $(g_V,m_V)$ yield the correct total relic abundance. When projected into the $(g_V,m_V)$ plane, the viable points trace out a characteristic ``S-curve'' relation.  

This relation plays a crucial role in our analysis: for each $(m_s,\sin^22\theta)$ point in the scan, only those $(g_V,m_V)$ pairs lying on the S-curve are physically relevant and are passed on to subsequent stages of our pipeline. In effect, this constraint reduces the dimensionality of the model parameter space: rather than scanning a full four-dimensional space $(m_s,\sin^22\theta,g_V,m_V)$, the problem is effectively reduced to three dimensions.  

Operationally, we determine the S-curve for each $(m_s,\sin^22\theta)$ point by employing a numerical root-finding procedure. The root solver is coupled directly to the production code, iteratively adjusting $(g_V,m_V)$ until the calculated sterile neutrino abundance matches the observed dark matter density. The resulting set of viable $(g_V,m_V)$ points thus defines the surface of solutions that maps directly onto our subsequent cosmological and astrophysical analyses. In practice, there are three regions along the curve which have distinct phenomenological characteristics. 

\begin{figure}[t!]
    \centering
    \includegraphics[width=\columnwidth]{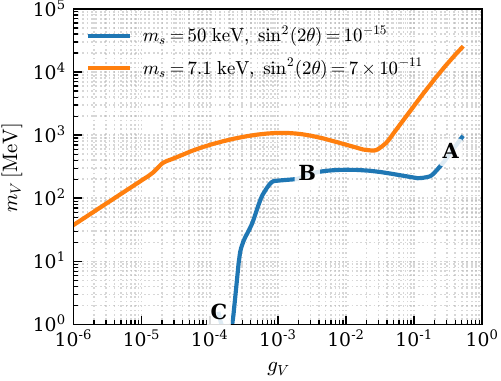}
    \caption{The $g_V$-$m_V$ relations are presented for two points using the Neutrinophilic production model. Each curve represents the valid region of parameter space where the production creates the full relic dark matter density. Point A highlights the high-mass case, point B highlights the intermediate mass case, while point C highlights the low mass case.} 
    \label{fig:s-curve}
\end{figure}

\begin{figure}[t!]
    \centering
    \includegraphics[width=\columnwidth]{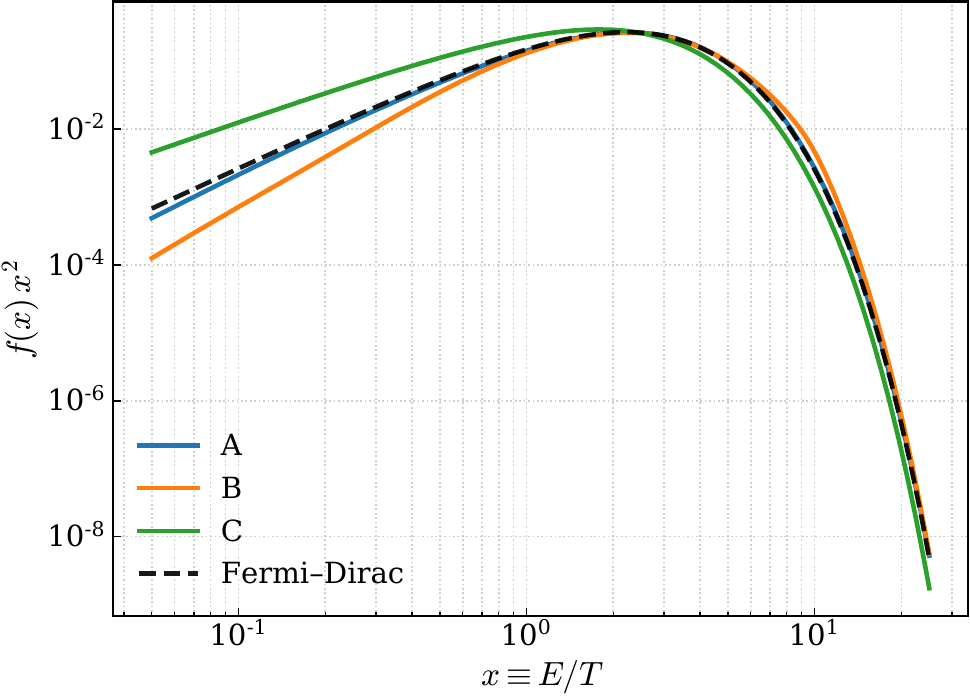}
    \caption{Shown here are the PSDs associated with the labeled points in Fig.~\ref{fig:s-curve}.} 
    \label{fig:PSD}
\end{figure}

As an example, Fig.~\ref{fig:s-curve} shows this characteristic relation for two example points, labeled within the figure. The PSDs associated with them are shown in Fig.~\ref{fig:PSD}.

Further details of the apparent structure are given in Sec.~\ref{subsec:modes}.

\subsection{Mapping to structure formation}\label{subsec:structure}

For each PSD output at $T_{\mathrm{low}}$ we compute the linear matter power spectrum with the Cosmic Linear Anisotropy Solving System (CLASS)~\cite{Blas:2011rf}, using \emph{Planck} 2018 baseline parameters, specifically the best estimate parameters from the TT,TE,EE+lowE+lensing column in Table 2 and including one massive neutrino with $\sum m_\nu=0.06$ eV~\cite{Planck:2018vyg}. We then form the relative transfer function
\begin{equation}
T(k)\equiv\sqrt{P_{\rm model}(k)/P_{\Lambda\mathrm{CDM}}(k)}\,,
\end{equation}
and fit to the calibrated thermal-WDM template to define an \emph{equivalent} $\mth$ for each model point. We adopt the updated calibration of thermal-WDM suppression for large $\mth$ from~\cite{Vogel:2022odl}, which corrects the older Viel et al.\ mapping and is particularly important above a few keV.

\subsection{Momentum spectra and transfer functions}\label{subsec:modes}

Across the $(\gV,\mV)$ plane we identify three qualitative regimes, familiar from Ref.~\cite{Kelly:2020pcy}: (i) a contact-like heavy-mediator limit ($\mV\!\gg\!T$) where enhanced scattering dominates and the PSD tends to be relatively \emph{warm} (larger $\langle p/T\rangle$); (ii) an intermediate regime where dispersion and scattering both matter, yielding PSDs with colder cores and modest warm tails; and (iii) a light-mediator regime ($\mV\!\ll\!T$) with forward-scattering dominated potentials and screening-regulated rates that can efficiently populate low momentum modes, producing \emph{cooler} spectra. In all regimes we find that the linear-theory transfer function is well captured by a single-parameter thermal-WDM fit at the $\lesssim\!{\cal O}(10\%)$ level near the $T(k)=1/2$ scale, enabling a sufficiently accurate $\mth$ assignment model-by-model.

\subsection{Astrophysical and particle constraints used}\label{subsec:constraints}

Once mapped to $\mth$, each model point can be directly confronted with the most stringent astrophysical structure-formation probes. Joint analyses of Milky Way satellite counts and strong-lensing substructure require $\mth \gtrsim 8\,\mathrm{keV}$ at 95\%~C.L. \cite{Nadler:2021dft,Zelko:2022tgf}. Independent constraints from the one-dimensional Lyman-$\alpha$ forest flux power spectrum yield $\mth > 3.1\,\mathrm{keV}$ (95\%~C.L.) and show a mild preference for $\mth \simeq 4$--$5\,\mathrm{keV}$ \cite{Villasenor:2022aiy}. 

In parallel, we impose X-ray decay limits in the $(\ms,\sintwotheta)$ plane, drawing on Chandra, XMM-Newton, NuSTAR Galactic halo long-exposure and all-sky stray-light analyses together with stacked-field observations from INTEGRAL/SPI \cite{Perez:2016tcq, Roach:2019ctw, Roach:2022lgo, Krivonos:2024yvm, Fischer:2022pse}.

In addition to the above astrophysical sources of insight into the sterile neutrino parameter space, in the presence of an NSI mechanism, relevant particle physics data can be used to further constrain the model. These experimental data sources are numerous and reviewed in more detail in Ref.~\cite{Kelly:2020pcy}. In practice, we can apply a filter that removes all points in our 4D space that are excluded based on their $g_V$ and $m_V$ before using constraints from structure. It is only by taking these in combination that the most robust constraints can be applied to each NSI production mechanism for sterile neutrino dark matter. 


\begin{figure*}[t!]
    \centering
    
    \begin{minipage}{0.48\textwidth}
        \centering
        \includegraphics[width=\textwidth]{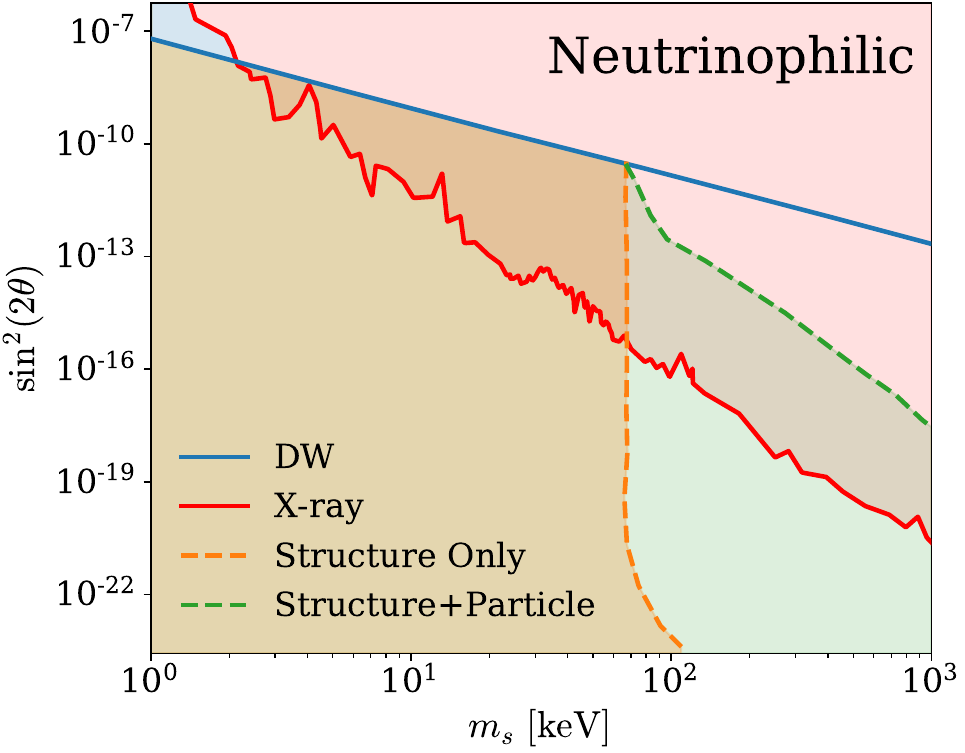}
    \end{minipage}\hfill
    \begin{minipage}{0.48\textwidth}
        \centering
        \includegraphics[width=\textwidth]{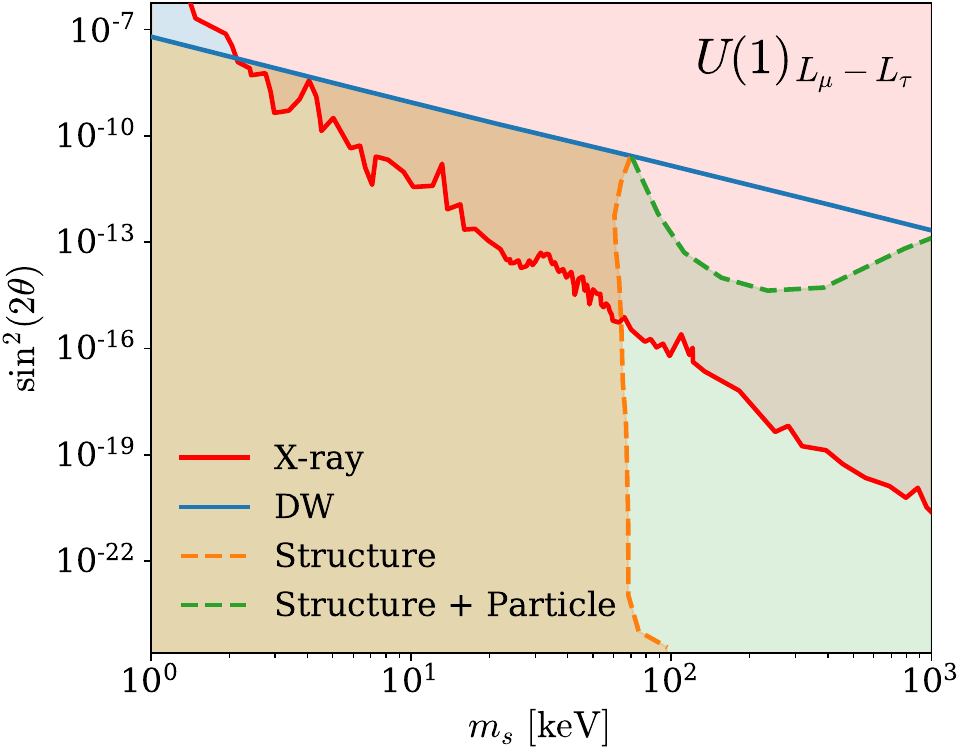}
    \end{minipage}

    \vspace{1em}
    \begin{minipage}{0.48\textwidth}
        \centering
        \includegraphics[width=\textwidth]{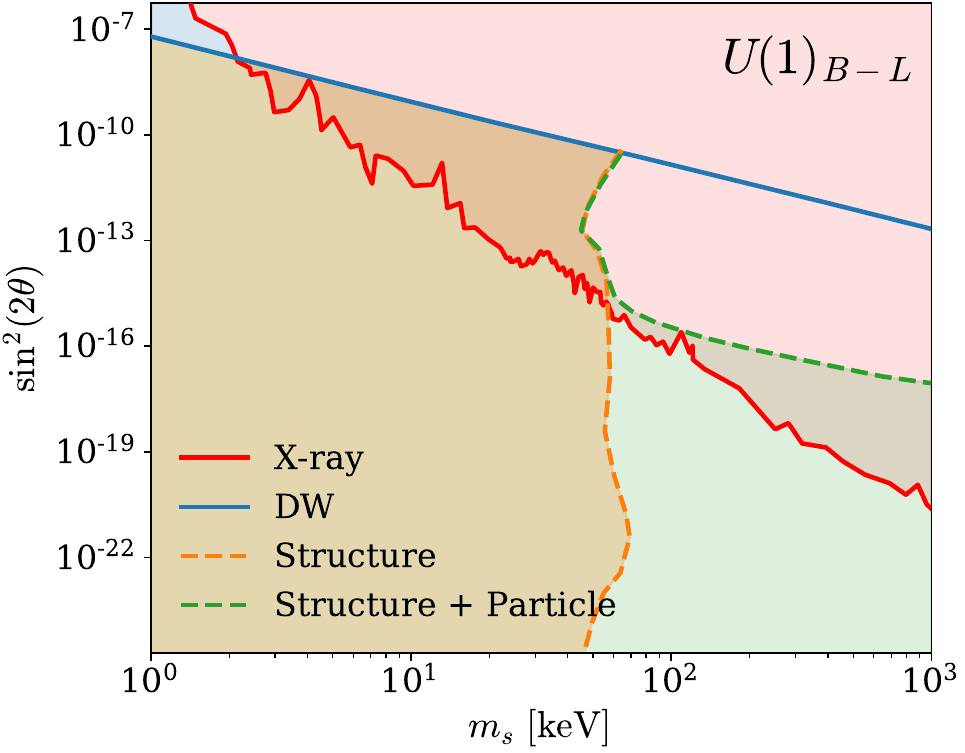}
    \end{minipage}\hfill
    \begin{minipage}{0.48\textwidth}
        \caption{Shown is the full sterile neutrino dark matter parameter space under Neutrinophilic vector-mediated NSI production, $L_\mu-L_\tau$ vector-mediated NSI production, and ${L_\mu-L_\tau}$ vector-mediated NSI production. The blue line is the standard DW production band. The dashed orange contour is the thermal WDM particle mass lower bound from the Zelko+~\cite{Zelko:2022tgf}, disfavoring the region left of the line. The green dashed line is structure in addition to particle physics bounds as described in Ref.~\cite{Kelly:2020pcy}. The red cross-hatched X-ray exclusion region is the combination of \textit{Chandra}, \textit{XMM-Newton}, \textit{NuSTAR} and \textit{SPI}/\textit{INTEGRAL} at 95\%~C.L. upper limits \cite{Perez:2016tcq, Roach:2019ctw, Roach:2022lgo, Krivonos:2024yvm, Fischer:2022pse}.}
        \label{fig:results}
    \end{minipage}
    \vskip -0.2 cm
\end{figure*}

\section{Results}
\label{sec:results}

In this section we present the main results of our analysis, focusing on three representative realizations of vector-mediated nonstandard interactions. For each case, we compare the constraints from structure-formation alone with those obtained after incorporating particle physics limits on the underlying interaction model. Figure~\ref{fig:results} shows the allowed regions in the $(\ms,\sintwotheta)$ plane, together with the impact of the additional model-specific constraints.

\subsection{Neutrinophilic}
\label{subsec:neutrinophilic_results}

For the neutrinophilic vector, production is efficient over a broad range of $(\ms,\sintwotheta)$ values. Figure~\ref{fig:results} shows the application of the various constraints to the full sterile neutrino parameter space. Using structure-formation data alone, the neutrinophilic scenario is excluded up to $\ms \simeq 70~\mathrm{keV}$. When combined with laboratory and astrophysical limits on the new interaction, however, the full parameter space is excluded at the 95\%~C.L. This demonstrates the complementarity of cosmological and particle physics probes.

\subsection{Gauged $U(1)_{L_\mu-L_\tau}$}
\label{subsec:LmuLtau_results}

In the $L_\mu - L_\tau$ case, the interplay between resonant production and stronger experimental limits leads to a more restricted parameter space. Figure~\ref{fig:results} displays the corresponding constraints mapped onto the full sterile neutrino parameter space. While the contour arising from cosmic structure is broadly similar to that seen in the neutrinophilic case, the stricter constraints coming from laboratory experiments, particularly from neutrino trident scattering with Columbia–Chicago–Fermilab–Rochester \cite{Altmannshofer:2014pba} lead to a very narrow region which bounds from both sides to create a small pocket of allowed parameter space after structure is applied on top of the particle constraints. The final result is the strong exclusion of the mechanism at much higher than 95\% ~C.L.

\subsection{Gauged $U(1)_{B-L}$}
\label{subsec:BL_results}

Finally, for the $B-L$ extension, the universal couplings to baryons and leptons lead to strong laboratory bounds, but the specific form of PSD produced tends to relax the combined particle-astrophysical bounds. The full sterile neutrino parameter space is shown in Fig.~\ref{fig:results}. We find that the $B-L$ model is still excluded, but only at the 95\%~C.L, particularly near the $\ms \simeq 70~\mathrm{keV}$ region. The relatively weak exclusion in this model results from the comparatively large mediator masses required for a given mixing angle as the sterile mass increases. We find that the coupling needed as sterile mass increases is not monotonic, instead increasing and then decreasing across the range of sterile neutrino masses considered. This feature is not sufficient to evade current constraints, and the production mechanism is unable to satisfy all constraints while producing the full dark matter density.

\section{Discussion and Conclusions}\label{sec:conclusions}

We developed and applied a production--to--structure pipeline for sterile-neutrino dark matter produced via vector-mediated nonstandard neutrino interactions. The framework extends DW production to include NSI thermal potentials and collision integrals with a validated light/heavy-mediator interpolation, generates nonthermal PSDs across $(\ms,\sintwotheta,\gV,\mV)$ grids, and maps each point to an equivalent thermal relic mass $\mth$ via transfer-function fitting in CLASS, while noting particle physics relevant parameters for use in additional screening. This enables appropriate use of structure-formation bounds for each model, alongside orthogonal X-ray decay constraints in the $(\ms,\sintwotheta)$ plane.

In the case of modifications to early-Universe cosmic history with low-reheating temperature scenarios, the lack of a sustained period of neutrino interactions above Big Bang nucleosynthesis allows for NSI production with larger active-sterile mixing, and commensurately smaller NSI strength \cite{Chichiri:2021wvw}. The constraints on the model parameters are relaxed in these scenarios. In our results, the approximate lower-bound on $m_s$ imposed by structure for all three vector-mediated models is at the $\sim\! 70 \; \mathrm{keV}$ scale. When combined with the results in Fig.~5 of Ref.~\cite{Chichiri:2021wvw}, this potentially allows for viable remaining parameter space. Exploration of low-reheating temperature NSI models' impact on structure formation is left for future work. 

In summary, our main finding is the exclusion of all three vector-mediated models studied at the 95\%~C.L. or above in the case of standard thermal history in cosmology. The sterile neutrino particle mass and mixing angle parameter space constraints on all three vector-mediated cases are very similar to those for the scalar mediated case when we only consider structure. However, the greater sensitivity of non-cosmological probes of vector-mediated models leads to these vector-mediated models being fully ruled out, while the scalar-mediated models remain viable. This result is significant in that it demonstrates the limitations of common NSI models that were previously understood to evade the wide variety of constraints imposed on sterile neutrino models seeking to produce the full dark matter density. While other NSI models, including scalar mediated models, remain viable in particular regimes \cite{An:2023mkf}, our framework provides an extended method for the application of the full space of constraints to the standard sterile neutrino parameter space. If dark matter is in the neutrino sector as a sterile neutrino, NSI in the active neutrinos produces a rich phenomenology that can be tested in the laboratory, in X-ray astronomy, as well as through structure formation, and it remains uncertain if or when a definitive signal will emerge.

\acknowledgements
CMV, HGE, and KNA acknowledge useful conversations with Rui An, Francis-Yan Cyr-Racine, Andr\'e de Gouv\^ea, George Fuller, Kevin Kelly, Jay Krishnan, Harri Parkkinen, Michael Ryan, Manibrata Sen, Brian Shuve, M.\ Cristina Volpe, Yue Zhang. We thank an anonymous referee for improving the paper. KNA is partially supported by U.S. National Science Foundation (NSF) Theoretical Physics Program Grant No.\ PHY-2210283. KNA acknowledges support of the Institut Henri Poincaré (UAR 839 CNRS-Sorbonne Université), and LabEx CARMIN (ANR-10-LABX-59-01) in hosting the ``Dark Matter and Neutrinos'' program, where many of these discussions took place. KNA thanks the Institute for Nuclear Theory at the University of Washington for its kind hospitality in hosting the INT-20-1a Neutrinos from the Lab to the Cosmos Program, which facilitated initial discussions regarding structure formation constraints on NSI-assisted sterile neutrino dark matter production models. This program and research was supported in part by the INT's U.S. Department of Energy grant No. DE-FG02-00ER41132.

\bibliography{NSI}
\bibliographystyle{apsrev4-1}

\end{document}